\documentstyle[11pt,newpasp,twoside,epsf]{article}
\reversemarginpar

\begin{document}
\title{Puzzling Pulsars and Supernova Remnants}
\author{D.R.~Lorimer}
\affil{Arecibo Observatory, HC3 Box 53995, Arecibo, PR 00612, USA}
\author{R.~Ramachandran}
\affil{NFRA, Dwingeloo \& Astronomical Institute, University of Amsterdam}

\begin{abstract}
The fact that the majority of the youngest radio pulsars are
surrounded by expanding supernova remnants is strong evidence that
neutron stars are produced in the supernovae of massive stars. In many
cases, the pulsar appears significantly offset from the geometric
centre of the supernova remnant, indicating that the neutron star has
moved away from the site of the explosion with a substantial space
velocity since birth. Here we show that the
these offsets show an overwhelming preference for one sign in terms of
Galactic longitude, a result that has important implications
for the number of genuine associations. The origin of this statistically
significant effect may lie in a differential Galactic rotational velocity
between stars and gas in the interstellar medium.
\end{abstract}

\noindent
An outstanding question in pulsar statistics is what fraction of
proposed pulsar-supernova remnant associations are genuine as opposed
to chance line-of-sight alignments. In attempting to answer this
question, Gaensler \& Johnston (1995) noted that the
distribution of positional offsets shows a clear excess at small
values compared to that expected by chance. These analyses suggest
that between 7 and 17 of the 30 presently proposed pairs are genuine
(Lorimer et al.~1998). Resolving the offsets into Galactic coordinates
we find an important and unexpected twist in this story.

The positional data for 27 pulsar-supernova remnant pairings presented in
Fig.~1 are the result of a cross-correlation of the publicly-available
pulsar and supernova remnant catalogues (Taylor et al.~1995; Green 1998). 
For each pairing we computed the positional offset defined as the difference
between the position of the pulsar and the geometric centre of the
remnant. Following Shull et al.~(1989) we normalised
each offset to the angular size of the respective remnant. Although
pulsar positions are known to within an arc second from timing
measurements, positions of remnant centroids are more difficult to
determine. We carefully searched the available literature
to check the positions listed in the catalogue and obtain conservative
estimates of their uncertainty.

It is immediately apparent that the distribution of offsets shown in
Fig.~1 is not uniform. This is easily quantified by counting the
number of pairs $N$ within a given angular radius $r$. For a uniform
distribution, $N(<r)\propto r^2$ so that the number expected within $2r$
should be $4N$.  This is clearly not the case, with 10 and 18 pairs
lying within half and one remnant radii respectively. What is 
most remarkable is the {\it sign} of the offsets. Whilst the signs in
Galactic latitude show no preference for positive or negative values,
we see that 20 out of the 27 pairs have negative offsets in Galactic
longitude. A simple binomial calculation suggests that
this has only a 0.7\% likelihood of happening by chance. 
All the 10 pulsars that lie within half a remnant radii have
negative longitude offsets.

\begin{figure}[hbt]
\setlength{\unitlength}{1in}
\begin{picture}(0,3.2)
\put(0.5,4){\includegraphics{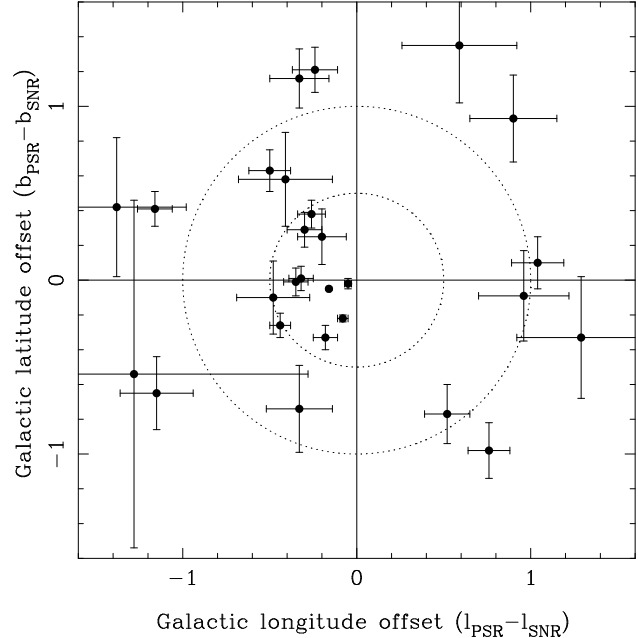}}
\end{picture}
\caption{
The distribution of normalised position offsets for pulsar-supernova
remnant pairs. All offsets have been normalised to the angular size of
the parent remnant. The normalisation allows us to display all the
offsets relative to the size of the remnant shell (the circle of unit
radius). Characterising the pairs in this way is extremely powerful
since it does not rely on the distances to individual pulsars or
supernova remnants, quantities that are often difficult to obtain
reliably.
}
\end{figure}

We have no reason to suspect that this highly significant effect is a
result of measurement uncertainties which do not explain the absence
of any trends in the latitude offset distribution. 
We conclude that the data are showing an intrinsic effect common to
both the pulsars and supernova remnants.  Although the physical origin
of this effect is unclear, one possibility is that it is due to a
differential velocity between stars and gas in the interstellar
medium. We are presently investigating this more carefully using
Monte Carlo simulations.

\end{document}